\documentclass[conference]{IEEEtran}
\usepackage{romannum}
\usepackage{amsmath}
\usepackage{graphicx}
\usepackage{multirow}
\usepackage{amssymb,bm}
\usepackage{algpseudocode}
\usepackage{algorithm}
\usepackage{array}
\usepackage{subfigure}
\usepackage{color}
\usepackage[utf8]{inputenc}
\usepackage[T1]{fontenc}
\usepackage{url}
\usepackage{ifthen}
\usepackage{cite}
\usepackage{mathtools}

\ifCLASSINFOpdf
\else

\fi

\begin{document}

\title{Deep Learning Assisted Sum-Product Detection Algorithm for Faster-than-Nyquist Signaling}

\author{%
  \IEEEauthorblockN{Bryan Liu, Shuangyang Li, Yixuan Xie and Jinhong Yuan\\}
  \IEEEauthorblockA{University of New South Wales, Sydney, NSW, Australia\\
                    Email: \{bryan.liu, shuangyang.li, yixuan.xie, j.yuan@unsw.edu.au\}}
}

\maketitle

\begin{abstract}
A deep learning assisted sum-product detection algorithm (DL-SPA) for faster-than-Nyquist (FTN) signaling is proposed in this paper. The proposed detection algorithm concatenates a neural network to the variable nodes of the conventional factor graph of the FTN system to help the detector converge to the \emph{a posterior} probabilities based on the received sequence. More specifically,
the neural network performs as a function node in the modified factor graph to deal with the residual intersymbol interference (ISI) that is not modeled by the conventional detector with a limited number of ISI taps.
We modify the updating rule in the conventional sum-product algorithm so that the neural network assisted detector can be complemented to a Turbo equalization. Furthermore, a simplified convolutional neural network is employed as the neural network function node to enhance the detector's performance and the neural network needs a small number of batches to be trained. Simulation results have shown that the proposed DL-SPA achieves a performance gain up to $2.5$ dB with the same bit error rate compared to the conventional sum-product detection algorithm under the same ISI responses.
\end{abstract}
%\begin{IEEEkeywords}
%high-density parity-check codes, reliability-based decoding, deep learning.
%\end{IEEEkeywords}
\IEEEpeerreviewmaketitle

\section{Introduction}
With the growing demand of high speed data transmissions, faster-than-Nyquist (FTN) signaling \cite{FTN_Mazo,FTN} has recently regained its popularity.
Different from the conventional methods of enhancing the data rate, which normally requires more time/bandwidth/spatial resources, FTN signaling
enhances the spectral efficiency by intentionally transmitting the symbols faster than the Nyquist rate without increasing the bandwidth consumption.
Therefore, FTN signaling has been largely considered in different communication applications, such as satellite communications \cite{TFP}, and 5G or beyond 5G communications \cite{FTN_5G}.

A major drawback of FTN signaling is that the higher symbol rate induces inevitable and severe intersymbol interference (ISI) at the transmitter side, which requires a very complex detector at the receiver side \cite{FTN}. For example, the number of states in a BCJR detector increases exponentially with the constellation size and number of ISI taps.
For a coded FTN system, the Turbo equalization is usually applied at the receiver, where iterations are performed between the BCJR detector and channel decoder. The overall detection/decoding complexity further increases with respect to the number of iterations.
Therefore, designing practical reduced-complexity detectors is a major research topic for FTN signaling. Two M-BCJR algorithms for detecting FTN signaling were proposed in \cite{Shuangyang} based on the Ungerboeck observation model \cite{Ungerboeck}, and they show promising error performance for coded FTN systems by applying Turbo equalization.
A soft-in soft-out (SISO) detection algorithm was proposed in \cite{SISO}, where the sum-product algorithm is applied to a suitable factor graph (FG) based on the Ungerboeck observation model.
The complexity of the algorithm is linear in the number of interferers during each iteration in contrast to the BCJR algorithm.
However, reduced-complexity detection algorithms usually undermine the error performance.
For instance, there are mainly two different aspects that may contribute to the performance loss for the SISO algorithm in \cite{SISO}.
Firstly, since the number of ISI taps for FTN signaling is infinite in theory \cite{FTN}, if we only consider the most significant ISI taps in detection, the residual ISI taps will degrade the error performance of the system. Secondly, cycles contained in the FG may accumulate the correlation between the messages during the detection iterations and thus affect the error performance.
Therefore, to improve the performance of the existed reduced-complexity detection algorithms for FTN signaling, we consider to utilize a neural network to compensate the performance loss.

Recently, deep learning supplemented detection algorithms and decoding algorithms are explored by researchers to further enhance the performance of a communication system.
This includes the research of the autoencoders \cite{Autoencoder_1, Autoencoder_2} and the neural network optimization schemes which transform the FGs into neural network systems \cite{DLLC, BP_adapt}.
For the autoencoders, a neural network system with multiple layers is employed and trained to overcome the issues such as multipath interferences and signal distortions.
However, since the connection among the multiple layers of the neural network model does not rely on the mathematical models of the channels, the neural network usually needs a large number of training samples, generally more than $2^K$ \cite{DLSTB}, where $K$ is the information sequence length, to converge to a good performance.
On the other hand, the "unfolded" neural network detection or decoding algorithms take advantage of the well-developed channel models \cite{DLLC}, which lead to specific neural network connections. However, the flexibility of the neural network designs is neglected. Such a neural network can only optimize the performance based on a constant graph, which may not lead to the globally optimized performance.

In this paper, we propose a neural network approach to compensate the performance loss for the sum-product detection algorithm (SPDA) proposed in \cite{SISO} for detecting coded FTN signals.
We modify the FG of the SPDA by connecting an arbitrary neural network to the variable nodes (VNs) via additional function nodes (FNs), where the tunable parameters over the edges of the FG are optimized through training.
The inaccuracy of the messages over the edges of the FG are expected to be compensated by the neural network.
Furthermore, a new message updating rule is proposed so that the proposed deep learning assisted sum-product detection algorithm (DL-SPA) can be easily implemented in a Turbo equalization fashion without the need of optimization corresponding to any particular channel decoder.
The proposed algorithm maintains the flexility of neural network designs while reducing the number of training samples by taking advantage of the modified FG.
Simulation results show that the proposed DL-SPA outperforms the SPDA under an FTN system with the same symbol rate, and it shows a better performance compared to the truncated BCJR algorithm \cite{TrunBCJR} with the same number of considered ISI taps.

%This paper is organized as follows. The next section will briefly describe a FTN system along with the Turbo equalization system. The sum-product detection algorithm proposed in \cite{SISO} is also briefly reviewed in the next section. In Section \Romannum{3}, we introduce the proposed DL-SPA accompanies with a new update rule and a simplified convolutional neural network. We discuss how the modification of the update rule to the messages to be conveyed to the neural network will improve the performance of the overall Turbo equalization system when the trained detector is employed. Simulation results are shown in Section \Romannum{4} and lastly, concluding remarks are delivered in Section \Romannum{5}.

%\vspace{-1mm}
\section{Preliminaries}
\subsection{Coded FTN and system model}
\begin{figure}[t!]
	\centering
	\includegraphics[width=90mm]{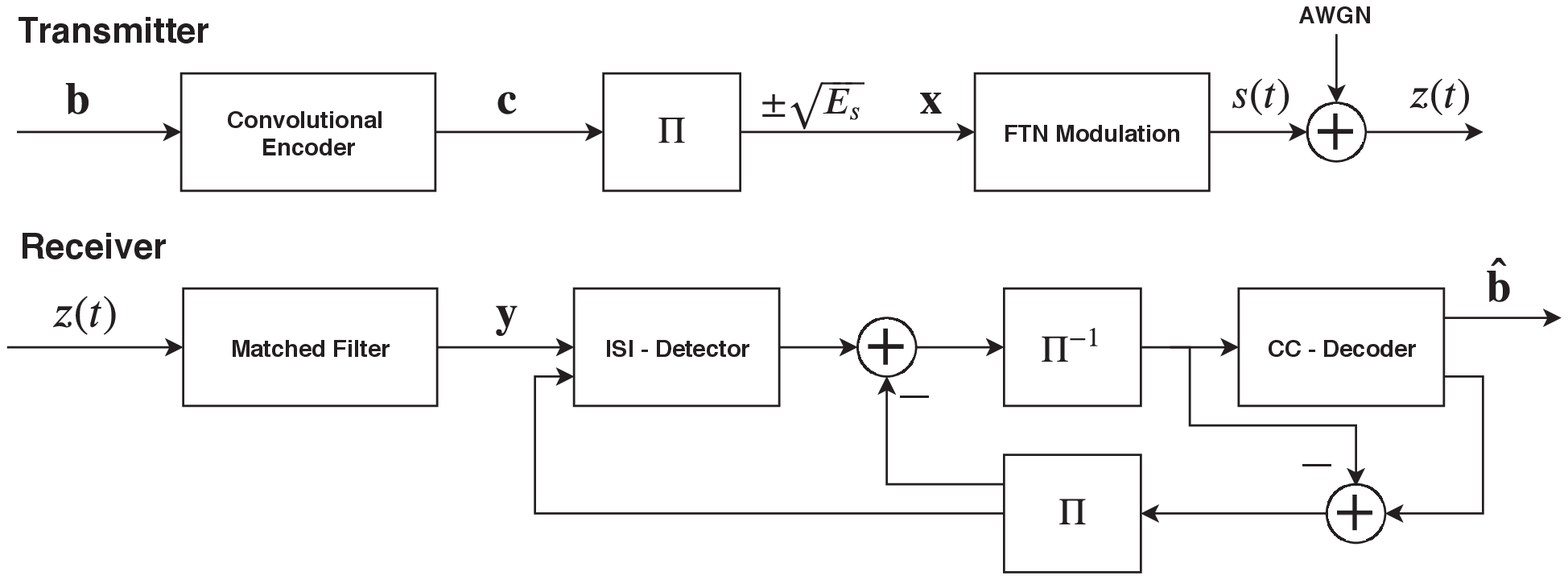}
    \vspace{-3mm}
	\caption{System model.}
    \label{TE}
\vspace{-3mm}
\end{figure}

Without loss of generality, a model of an FTN system is shown in Fig. \ref{TE}. Let $\bm{b}$ denote the source data with length $K$.
In the transmitter, $\bm{b}$ is convolutionally encoded, resulting in a codeword $\bm{c}$.
A sequence of $N$ binary phase-shift keying (BPSK) symbols $\mathbf{x} = [x_1, x_2, ..., x_N]^\text{T}$ is generated after interleaving the bits in $\bm{c}$.
FTN signals are linear modulation signals of the form $s(t) = \sqrt{E_s}\sum_n{x_n}{h(t-n\tau T)}$, where $\tau$ is the time acceleration factor of FTN signaling \cite{FTN_Mazo} and $h(t)$ is a $T$-orthogonal root raised cosine pulse with a roll-off factor $\alpha$.
Assume that the channel is corrupted by additive white Gaussian noise (AWGN) with a variance of $\sigma^2$.
The received sequence $\bm{y}$ after a matched filtering and FTN rate sampling is given by $\bm{y} = \mathbf{G}\mathbf{x} + \bm{\eta}$, where $\mathbf{G}$ is a Toeplitz generator matrix
constructed by the ISI taps ${g_i} = \int_{ - \infty }^\infty  {h\left( t \right){h^*}\left( {t - i\tau T} \right){\rm{d}}t}$, $|i| \le L$
and $\bm{\eta}$ has an autocorrelation matrix $\mathbb{E}[\bm{\eta} {\bm{\eta}}^H] = \sigma^2\mathbf{G}$.
Here, $L$ denotes the number of channel responses with significant energy, i.e., $|{{{g_k}} \mathord{\left/
 {\vphantom {{{g_l}} {{g_0}}}} \right.
 \kern-\nulldelimiterspace} {{g_0}}}| \ge  0.01, |k| \le L$. The rest ISI taps with insignificant energy are therefore negligible and then set to zeros for simplicity.
%\graphicspath{ {./resources/}}
Once the sequence $\bm{y}$ is observed, the receiver performs the Turbo equalization,
where the extrinsic information from the detector and decoder is exchanged iteratively via the interleaver $\Pi $ or deinterleaver ${\Pi ^{ - 1}}$
until the maximum iteration number is reached. The sequence \(\hat {\bm{b}}\) as the estimate of $\bm{b}$ is generated after the iteration, which is regarded as the output for the receiver.

\subsection{Sum-product detection algorithm}
A sum-product detection algorithm (SPDA) was proposed in \cite{SISO} based on the Ungerbock model \cite{Ungerboeck}. Given the received sequence $\bm{y}$, the SPDA algorithm factorizes the $a \ posterior$ probabilities (APPs) $P(\bm{x}|\bm{y})$ of the transmitted sequence $\bm{x}$ mainly based on three FNs:

$\bullet$ $O_i(x_i)$ for $i \in \{1,...N\}$: The $a \ priori$ probability that the symbol $x_i$ is transmitted.

$\bullet$ $T_i(x_i)$ for $i \in \{1,...N\}$: The symbol likelihood function that the symbol $x_i$ is transmitted based on the received symbol $y_i$.

$\bullet$ $I_{i,j}(x_i, x_j)$ for $i \in \{1,...N\}$ and $j \in \{1,...N\}$ that $i>j$: The FN that conveys the APPs from the interfering nodes.

The functions of $T_i(x_i)$ and $I_{i,j}(x_i, x_j)$ are defined as \cite{SISO}:
\vspace{-2mm}
\begin{gather}\label{T}
T_i(x_i) = \text{exp}\bigg{[}\frac{1}{\sigma^2}\text{Re}\bigg{\{}y_ix_i^*-\frac{{\bf{G}}_{i,i}}{2}|x_i|^2\bigg{\}}\bigg{]},\vspace{-5mm}
\vspace{-5mm}
\end{gather}
\vspace{-5mm}
\begin{gather}\label{I}
I_{i,j}(x_i, x_j) = \text{exp}\bigg{[}-\frac{1}{\sigma^2}\text{Re}\big{\{}{\bf{G}}_{i,j}x_ix_j^*\big{\}}\bigg{]},
\vspace{-5mm}
\end{gather}
where $x_i^*$ refers to the conjugate of the symbol $x_i$, $\text{Re}\{ \cdot \}$ represents the function that returns the real part of a value, and ${\bf{G}}_{i,j}=g_{i-j}$ is the ${(i-j)}$-th ISI tap. It is derived in \cite{SISO} that $P(\bm{x}|\bm{y}) \propto \prod_{i=1}^{N}\bigg{[}O_i(x_i)T_i(x_i)\prod_{j<i}I_{i,j}(x_i,x_j) \bigg{]}$.

%\graphicspath{ {./resources/}}
\begin{figure}[t!]
	\centering
	\includegraphics[width=50mm]{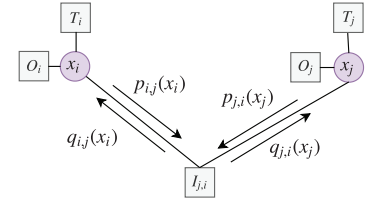}
    \vspace{-3mm}
	\caption{Message updating in the factor graph of SPDA.}
    \label{SPDA_Graph}
\vspace{-4mm}
\end{figure}

Define $q_{i,j}(x_i)$ as the message from the FN $I_{i,j}$ to the VN $x_i$, $p_{i,j}(x_i)$ as the message from the VN $x_i$ to the FN $I_{i,j}(x_i,x_j)$, $o(x_i)$ as the message from the VN $x_i$ to the FN $O_i$, and $Q_i(x_i)$ as the product of all messages incoming to the VN $x_i$, respectively. Here, $Q_i(x_i)$ indicates the proportional probability of the (approximated) APP $P(x_i|\bm{y})$ \cite{SISO}. Fig. (\ref{SPDA_Graph}) shows the messages to be updated between two VNs and the SPDA can be summarized as the updates of the following messages \cite{SISO}:
\begin{gather}\label{Q_update}
Q_i(x_i) = O_i(x_i)T_i(x_i)\prod_{j \neq i}{q_{i,j}(x_i)},
\vspace{-5mm}
\end{gather}
\vspace{-4mm}
\begin{gather}\label{o_update}
o_i(x_i) = \frac{Q_i(x_i)}{O_i(x_i)},
\vspace{-5mm}
\end{gather}
\vspace{-4mm}
\begin{gather}\label{p_update}
p_{i,j}(x_i) = \frac{Q_i(x_i)}{q_{i,j}(x_i)},
\vspace{-5mm}
\end{gather}
\vspace{-4mm}
\begin{gather}\label{q_update}
q_{i,j}(x_i) = \sum_{x_j}I_{i,j}(x_i, x_j)p_{j,i}(x_j)
\vspace{-5mm}
\end{gather}

The messages $\{p_{i,j}\}$ and $\{q_{i,j}\}$ are initialized to the same positive values \cite{SISO} and the messages are updated iteratively until the maximum number of iterations is reached.
The number of the FNs $I_{i,j}(x_i)$ linked to each VN $x_i$ increases linearly with the number of channel taps $L_E$ that are considered by the detector, where $L_E \le  L$. Once $O_i(x_i)$, $T_i(x_i)$, and $I_{i,j}(x_i, x_j)$ have been initialized, the SPDA conveys messages between the VNs and FNs to iteratively update the APPs of the transmitted symbols.

The SPDA computes the APPs by updating the messages from the FG. However, in practice, only $L_E$ taps are considered due to the detection complexity. Moreover, the FG contains short cycles which accumulate the correlation between the messages during the iterative update and will affect the algorithm's performance, especially for a coded FTN system with short codeword length. Therefore, we propose the DL-SPA to enhance the detection's performance for a coded FTN system.

\section{Proposed DL-SPA algorithm}
In this section, we introduce the proposed deep learning assisted sum-product algorithm (DL-SPA).
We propose to add a neural network to the original FG proposed in \cite{SISO}, where the neural network is linked to the VNs of the FG via additional neuron FNs.
The neural network is expected to convey the information from the residual ISI responses which are not considered by the original FG, and reduces the correlation accumulated in the cycles of the FG. We show how the messages to be passed to the neural network are modified to make the DL-SPA suitable for Turbo equalization.
One feature of our proposed structure is that we can train the DL-SPA without the prior knowledge of the decoder.
Therefore, once the neural network is trained, the extrinsic information from the decoder can be passed to the ISI-detector without further tuning the parameters in the neural network.
Furthermore, we propose a convolutional neural network (CNN) structure for DL-SPA for the sake of a simple training process. It also suits the detection algorithm due to the special convolutional property.

\subsection{New FG model and modified message updating rule}
\begin{figure}[t!]
	\centering
	\includegraphics[width=90mm]{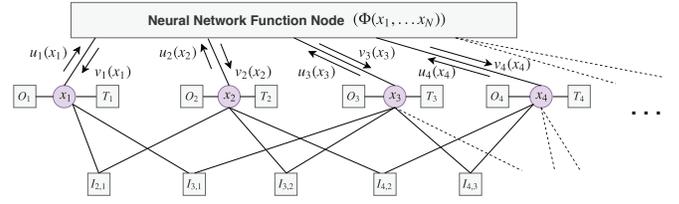}
    \vspace{-2mm}
	\caption{Deep learning assisted sum-product algorithm with $L_E$ = 2.}
    \vspace{-2mm}
    \label{DL-SPA}
\vspace{-4mm}
\end{figure}

Note that conventional neural network assisted detection or decoding algorithms introduce weights to the FG then unfold the message-passing algorithm to a neural network system for training and optimization \cite{DLLC}. In the SPDA, trainable weights can be attached to the messages $p_{i,j}(x_i)$, so the update rule in (\ref{q_update}) is modified into:

\vspace{-3mm}
\begin{gather}\label{q_update_new}
q_{i,j}(x_i) = \sum_{x_j}I_{i,j}(x_i, x_j)w_{j,i}p_{j,i}(x_j),
\vspace{-3mm}
\end{gather}
where $w_{j,i}$ is the weight attached to the message $p_{j,i}(x_j)$. Training with the additional weights improves the message passing algorithm's performance in a high SNR region \cite{DLLC}.
However, since the same model or connection is employed in the FG and the conventional neural network, the performance improvement by attaching trainable weights to the neural network is limited. In this work, we propose to concatenate a neural network FN $\Phi(x_1,...,x_N)$ to the VNs in the FG to compensate the effects of the residual ISI responses and the correlation induced along the short cycles.
As shown in Fig. \ref{DL-SPA}, different from the traditional FG, we nest a neural network to the VNs $x_i$ of the FG, for $i \in \{1,...,N\}$.
There are mainly two aims of nesting a neural network to the FG:

$\bullet$ The neural network connects to all the VNs. It is expected that all ISIs among the VNs are considered by the neural network. This is simpler compared to the FG that considers all ISI taps on one VN.

$\bullet$ The correlation induced during the iteration is expected to be compensated by the neural network. The APPs computation for all the VNs in each iteration can be optimized after tuning the parameters in the neural network.

Define $u_i(x_i)$ as the message from the variable node $x_i$ to the FN $\Phi(x_1,...,x_N)$ and $v_i(x_i)$ as the message from the FN $\Phi(x_1,...,x_N)$ to the variable node $x_i$.
The conventional sum-product algorithm sums all the intrinsic information for each variable node before passing the extrinsic information to the FN for further processing. This indicates that in a conventional sum-product algorithm, $u_i(x_i)=O_i(x_i)T_i(x_i)\prod_{j \neq i}{q_{i,j}(x_i)}$.
However, in a Turbo equalization, the extrinsic information from the decoder will pass to the detector.
Optimizing a Turbo equalization with a neural network system leads to two major problems.
Firstly, the training complexity will be largely increased if the decoder is also covered by the neural network layers.
Secondly, the optimization of the neural network needs to consider the specific channel decoder, which is inflexible from the design perspective, and undermines the generality of the ISI detector.

%\graphicspath{ {./resources/}}

To train an ISI detector which is applicable to a Turbo equalization without the prior knowledge of the decoder, we propose to only pass the soft information from the FN $I_{i,j}(x_i,x_j)$ to the neural network. This indicates that the message $u_i(x_i)$ will be updated by:

\vspace{-3mm}
\begin{gather}\label{u_update}
u_{i}(x_i) = \prod_{j \neq i}q_{i,j}(x_i).
\vspace{-3mm}
\end{gather}
The final \emph{a posterior} probability $Q_i(x_i)$ update rule becomes:
\begin{gather}\label{Q_update_new}
Q_i(x_i) = O_i(x_i)T_i(x_i)v_i(x_i)\prod_{j \neq i}{q_{i,j}(x_i)}.
\vspace{-3mm}
\end{gather}
Compared with the APP update rule in Eq. (\ref{Q_update}), Eq. (\ref{Q_update_new}) contains the message $v_i(x_i)$ from the neural network to the VN $x_i$.

Since in each iteration of the Turbo equalization, the value of $T_i(x_i)$ is constant, the messages from $T_i(x_i)$ will not be passed to the neural network for further processing.
The messages conveyed from the FN $I_{i,j}(x_i,x_j)$ contain correlations due to the short cycles in the FG. Besides reducing the training complexity by passing only the messages from the FN $I_{i,j}(x_i,x_j)$, the APPs accumulated at each iteration of the detection algorithm are tuned by the neural network's output messages $v_i(x_i)$. The DL-SPA can be summarized as follows:
\begin{enumerate}
  \item Update all the $a \ posterior$ probabilities $\{Q_i\}$ as in (\ref{Q_update_new});
  \item Update the messages $\{p_{i,j}\}$ as in (\ref{p_update});
  \item Update the messages $\{q_{i,j}\}$ as in (\ref{q_update});
  \item Update the messages $\{u_i\}$ as in (\ref{u_update});
  \item Compute the messages $\{v_i\}$ based on the trained $\Phi(x_1,...,x_n)$;
  \item If the maximum number of iterations is not reached, then go back to step 1;
  \item Update all the $a \ posterior$ probabilities $\{Q_i\}$ as in (\ref{Q_update_new}).
\end{enumerate}

\subsection{DL-SPA with simplified convolutional neural network and its training procedure}
We propose to employ a simplified convolutional neural network (CNN) assisted SPDA by considering the special convolution structure of CNN \cite{CNN}.
CNNs are widely used in image recognition systems.
Traditional CNNs usually involve several convolutional layers (Conv) and max-pooling layers.
The convolutional layer performs the convolution operation of the filters.
The filters convolve and stride over the input. The max-pooling layer performs downsampling to reduce the spatial size of the convolved features.
A dense layer is appended after the max-pooling layer to provide possibly nonlinear function \cite{CNN}. In \cite{CNNDetection}, a pure CNN based detection algorithm was proposed, where both max-pooling layers and dense layer are removed to reduce the training complexity, but a large number of convolutional layers and filters are kept.
In this paper, we remove the max-pooling layer and simplify the convolutional neural network to only have one convolutional layer. The message $u_i(x_i)$ contains two values of probabilities due to the BPSK modulation. The convolutional layer has $f$ filters and each filter has a size of $2 \times \kappa$, where 2 refers to the size of constellations and $\kappa$ indicates that $\kappa$ adjacent messages of $u_i(x_i)$ are considered by the filter.
The stride of the filter in the convolutional layer is set to be 1.
The filter convolves with the VNs from $u_1(x_1)$ to $u_N(x_N)$. This indicates that all the VNs are processed by the CNN FN. The output of the convolutional layer is reshaped to the same dimension as the input, then sent to a dense layer. The dense layer processes all the filters' results then output the message $v_i(x_i)$ for $i \in \{1,...N\}$.
A rectified linear activation function (ReLU) is used for both the convolutional layer and the dense layer \cite{ReLU}. The CNN has initial weights and biases randomly generated. Every iteration, messages $u_i(x_i)$ will be updated by Eq. (\ref{u_update}) then passed to the CNN. Messages are sent back from the CNN to join the APPs accumulation according to Eq. (\ref{Q_update_new}).
By ``unfolding'' the iterative message-passing algorithm to a NN system with multiple layers, the overall detection performance can be trained and optimized.

%\graphicspath{ {./resources/}}
\begin{figure}[t!]
	\centering
	\includegraphics[width=60mm]{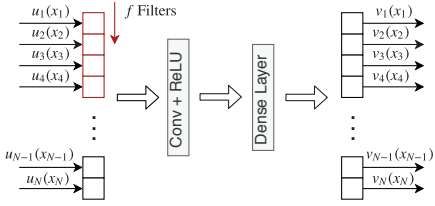}
    \vspace{-5mm}
	\caption{Simplified CNN function node $\Phi(x_1,...,x_N)$.}
    \label{CNN}
\vspace{-5mm}
\end{figure}

Equation (\ref{Q_update_new}) computes the APP of the transmitted symbol $x_i$. A sequence of log-likelihood ratios (LLRs) can further be acquired for every iteration of DL-SPA, where the LLR($x_i$)=$\text{log}\frac{P(x_i=+1|\bm{y})}{P(x_i=-1|\bm{y})}$. This allows us to setup a multi-loss function as introduced in \cite{DLLC} and \cite{Syndrome_based} to train the tunable parameters in the FG, which include the weights in Eq. (\ref{q_update_new}) and the weights and biases introduced in the CNN. Define $\hat{R}_i^m(x_i)$ as the LLR of the VN $x_i$ at the $m$-th iteration and $m_{max}$ as the maximum number of iterations of DL-SPA. Let the cross-entropy function be $\mathcal{F}_{ce}(R, \hat{R}^m) = -\frac{1}{N}\sum_{i=1}^{N}\big{(}R_{x_i}\text{log}(\frac{1}{1+e^{-\hat{R}^m_{x_i}}}) + (1-R_{x_i})\text{log}(1-\frac{1}{1+e^{-\hat{R}^m_{x_i}}})\big{)}$, where $R$ is the ground-truth label of the transmitted bits. The final loss function for training the neural network is given by:

\vspace{-5mm}
\begin{gather}\label{loss_func}
\Lambda = \sum_{m=1}^{m_{max}}\gamma^{m_{max}-m}\mathcal{F}_{ce}(R, \hat{R}^m),
\vspace{-4mm}
\end{gather}
where $\gamma < 1$ is a discount factor to adjust the loss at each iteration. During the training phase, a batch of random transmitted symbols over a range of signal-to-noise ratios (SNR) is generated as samples to train the neural network.
Once the neural network is fully trained, the DL-SPA can be attached to the Turbo equalization to perform as an ISI detector to exchange the extrinsic information with the decoder.

\vspace{-1mm}
\section{Numerical Results}
In this section, we evaluate the performance of the proposed DL-SPA scheme over convolutional coded FTN systems.
Without loss of generality, we consider coded FTN systems
with $\tau=0.5$ and $\tau=0.6$, where the channel code is the terminated (7, 5) 4-state rate-1/2 non-recursive convolutional code (CC).
The length of the data bits in both cases is $K=62$ and a BCJR decoder is employed to decode the CC. To get a fair comparison, we also perform the SPDA \cite{SISO} and truncated-BCJR detection algorithm \cite{TrunBCJR} with terminated ISI trellis. Note that, additional symbols need to be transmitted to terminate the ISI trellis, so that the overall spectral efficiency for the truncated-BCJR algorithm is slightly lower than that of the SPDA and DL-SPA.
The hyper-parameters to train the neural network system are shown in Table \ref{table:1}:

\vspace{-0mm}
\begin{table}[h!]
\centering
\begin{center}
 \begin{tabular}{c| c}
\hline
 Optimizer & Root Mean Square Propagation \\
 \hline
 Learning rate &  0.001 \\
  \hline
 Batch size & 360 \\
 \hline
 SNR range (dB) & (3, 8)  \\
 \hline
 Batch per SNR & 60 \\
 \hline
 $\gamma$ & 0.95 \\
 \hline
 $m_{max}$ & 15 \\
 \hline
 $f$ & 15 $(\tau=0.6)$, 20 ($\tau=0.5$) \\
 \hline
 Filter size & 2 $\times$ 4 \\
 \hline
\end{tabular}
\end{center}
\vspace{-2mm}
\caption{Hyper-parameters for the training of DL-SPA.}
\label{table:1}
\vspace{-4mm}
\end{table}
The initial values of the weights and biases in each iteration's CNN are randomly generated from a truncated normal distribution with a standard deviation of 0.03.

\begin{figure}[t!]
	\centering
	\includegraphics[width=66mm]{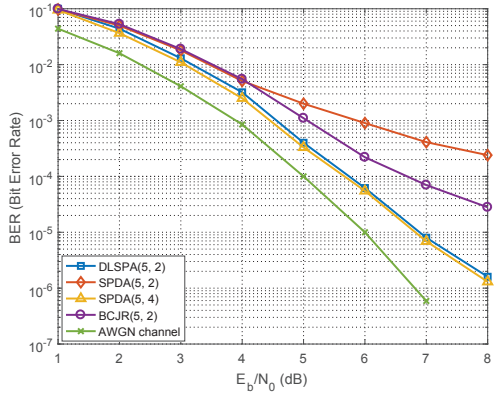}
    \vspace{-4mm}
	\caption{The BER of FTN signaling with $\tau = 0.6$, $\alpha=0.3$, CC(7,5).}
    \label{Tau06}
\vspace{-3mm}
\end{figure}

\begin{figure}[t!]
	\centering
	\includegraphics[width=66mm]{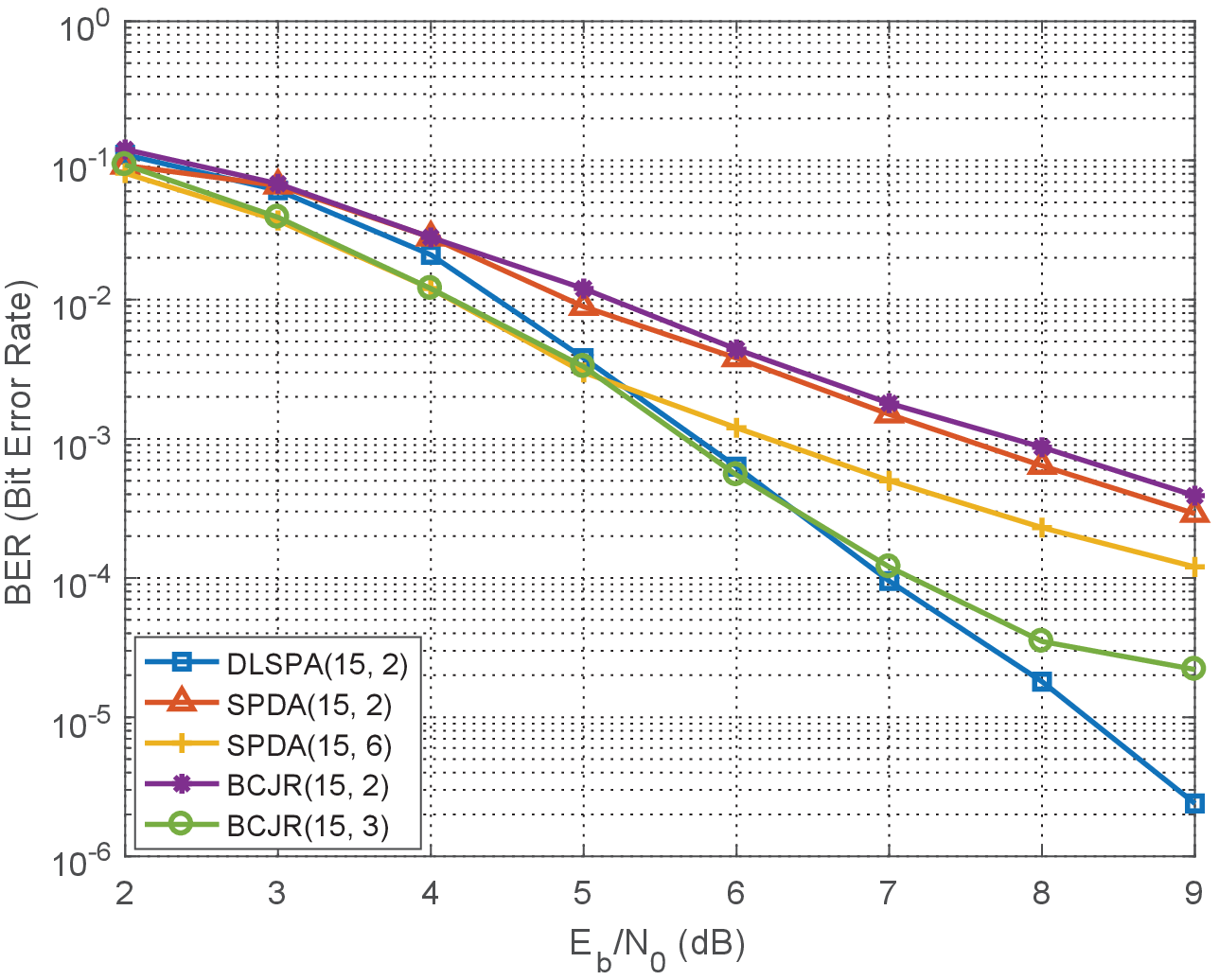}
    \vspace{-4mm}
	\caption{The BER of FTN signaling with $\tau = 0.5$, $\alpha=0.3$, CC(7,5).}
    \label{Tau05}
\vspace{-3mm}
\end{figure}

The bit error rate (BER) of various detection algorithms for the FTN systems is shown in Figs. \ref{Tau06} and \ref{Tau05}. Here, DL-SPA($\rho_{max}$, $L_E$) indicates the proposed DL-SPA detection method, and BCJR($\rho_{max}$, $L_E$) refers to the corresponding truncated BCJR detection algorithm \cite{TrunBCJR}. Here, $\rho_{max}$ indicates the number of iterations of the Turbo equalization. Both DL-SPA and SPDA utilize 15 iterations for updating the messages. It can be seen that for an FTN system with $\tau=0.6$, the SPDA has the worst error performance at a high SNR region, due to the lack of considerations of the residual ISI responses.
On the other hand, by adding the neural network FN, the DL-SPA(5, 2) can achieve similar error performance as the SPDA(5, 4).
In particular, the DL-SPA(5, 2) shows 2.5 dB and 0.5 dB gain to the SPDA(5, 2) and the BCJR(5, 2), respectively. The DL-SPA(5, 2) is 0.9 dB away from the CC decoding performance without ISI, i.e. the AWGN channel, which serves as the lower bound of the Turbo equalization.
For $\tau=0.5$, it can be seen that the DL-SPA(15, 2) outperforms the SPDA(15, 2) 1.8 dB at a BER=$10^{-3}$ and the DL-SPA(15, 2) has shown a 1.5 dB performance gain over the SPDA(15, 6) at a BER=$2 \times 10^{-3}$. The proposed DL-SPA(15,2) has shown a 0.4 dB gain at a BER=$3\times10^{-5}$ over the truncated BCJR(15,2). These results imply that the proposed algorithm outperforms the original SPDA and the BCJR algorithm with the same number of considered ISI responses, and reaches the performance of a more complex detection algorithm.

\begin{figure}[t!]
	\centering
	\includegraphics[width=66mm]{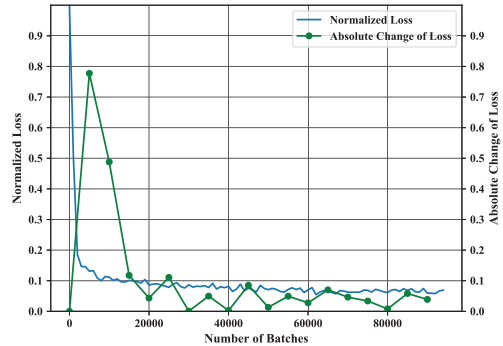}
    \vspace{-3mm}
	\caption{Normalized average loss for every $10^3$ batches of training samples and the absolute change of loss for every $5\times10^3$ batches.}
    \label{NL}
\vspace{-5mm}
\end{figure}

Fig. \ref{NL} illustrates the normalized average training loss for every $10^3$ training batches for FTN signaling with $\tau=0.6$. Define $\xi_{avg}^{a}$ as the average loss from $(a-1)\times5\times10^{3}$ to $a\times5 \times 10^{3}$ batches ($\xi_{avg}^{0}=1$) and $\xi_{cg}^a = |(\xi_{avg}^{a}-\xi_{avg}^{a-1})/\xi_{avg}^{a-1}|$ as the percentage of the absolute change on the average loss ($\xi_{cg}^{0}=0$), where $a \in \mathbb{Z}$. Define that a stable performance of the training is reached after $a\times5\times10^3$ batches, if $\xi_{cg}^{a'} < 0.1$ for any integer $a' > a$. From Fig. \ref{NL}, the training phase of the proposed algorithm takes roughly $360\times50000 = 1.8\times10^7$ samples to converge to a relative stable performance. Benefit from the derived FG and simplified neural network model, this number of training samples is much smaller than the conventional neural network decoders which generally need more than $2^{K}$ ($2^{62} \approx 4.6\times10^{18}$) training samples to converge to a good performance.

%\graphicspath{ {./resources/}}

%\graphicspath{ {./resources/}}

%\graphicspath{ {./resources/}}

\section{Conclusion}
In this paper, we proposed a deep learning assisted sum-product detection algorithm for FTN signaling.
By concatenating a NN to the FG of conventional FTN systems, the proposed detection algorithm computes the \emph{a posterior} probability with the help of the neural network.
A new message updating rule is proposed so that the proposed detection algorithm does not need to be optimized with respect to any particular channel decoder.
Furthermore, a simplified CNN architecture for the additional neural network FN is introduced to reduce the training complexity.
Simulation results show that the proposed DL-SPA provides a performance gain compared to the SPDA under the same number of ISI responses.
Meanwhile, benefiting from the well-developed model of the original factor graph and the simplified structure of the CNN, the DL-SPA needs a much smaller number of batches to train
the neural network compared to the conventional neural network decoder which needs at least $2^K$ training samples.

%\bibliographystyle{IEEEtran}
%\vspace{-1mm}
%\bibliography{ITW_refs}

\end{document}